\documentclass{aa}  
\usepackage{graphicx}
\usepackage{txfonts}
\begin{document}
\title{New evidence on the origin of the microquasar \object{GRO~J1655$-$40}}


\author{J.~A. Combi\inst{1}, L. Bronfman\inst{2}, I.F. Mirabel\inst{3,4}
          }
\authorrunning{Combi et~al.}
\titlerunning{New evidence on the origin of the MQ \object{GRO~J1655$-$40}}
\offprints{J.A. Combi}

\institute{Departamento de F\'{\i}sica (EPS), Universidad de Ja\'en,
Campus Las Lagunillas s/n, 23071 Ja\'en, Spain\\
\email{jcombi@ujaen.es}
\and
Departamento de Astronom\'{\i}a, Universidad de Chile, Casilla 36-D, Santiago, Chile\\
\email{leo@das.uchile.cl}
\and
European Southern Observatory, Alonso de Cordova 3107, Santiago, Chile
\and
On leave from CEA-Saclay, France\\
\email{fmirabel@eso.org}
             }

\date{Received; accepted }

 
  \abstract
   {}
   {Motivated by the new determination of the distance to the microquasar \object{GRO~J1655$-$40} by Foellmi et al. (2006), we conduct a detailed study of the distribution of the atomic (HI) gas, molecular (CO) gas, and dust around the open cluster NGC 6242, the possible birth place of the microquasar (Mirabel et al. 2002). The proximity and relative height of the cluster on the galactic disk provides a unique opportunity to study SNR evolution and its possible physical link with microquasar formation.}
   {We search in the interstellar atomic and molecular gas around NGC 6242 for traces that may have been left from a supernova explosion associated to the formation of the black hole in \object{GRO~J1655$-$40}. Furthermore, the 60/100 $\mu$m IR color is used as a tracer of shocked-heated dust.}
   {At the kinematical distance of the cluster the observations have revealed the existence of a HI hole of 1$^{\circ}$.5$\times$1$^{\circ}$.5 in diameter and compressed CO material acumulated along the south-eastern internal border of the HI cavity. In this same area, we found extended infrared emission with characteristics of shocked-heated dust. Based on the morphology and physical characteristics of the HI, CO and FIR emissions, we suggest that the cavity in the ISM was produced by a supernova explosion occured within NGC 6242. The lower limit to the kinematic energy transferred by the supernova shock to the surrounding interstellar medium is $\sim$ 10$^{49}$ erg and the atomic and molecular mass displaced to form the cavity of $\sim$ 16.500 $M_{\odot}$. The lower limit to the time elapsed since the SN explosion is $\sim$ 2.2$\times$10$^{5}$ yr, which is consistent with the time required by \object{GRO~J1655$-$40} to move from the cluster up to its present position. The observations suggest that \object{GRO~J1655$-$40} could have been born inside NGC 6242, being one of the nearest  microquasars known so far.}
{}


\maketitle
%

\section{Introduction}

Microquasars are X-ray binary systems exhibiting relativistic radio jets (Mirabel \& Rodr{\'{\i}}guez \cite{mira99}). These systems contain compact objects like stellar black holes or neutron stars that accrete matter from a mass donor companion star. The formation of stellar black holes may result from direct collapse of massive star progenitors or from delayed collapse of proto-neutron stars that undergo energetic explosions (Mirabel \cite{mira04}). In some cases, compact objects can suffer a strong natal kick imparted in the supernova event. Therefore, they can escape from the region of birth, where there could still exist signatures of their formation.      

Until now, only 15 microquasars are known in the galaxy (Paredes \cite{paredes05}). Among them, \object{GRO~J1655$-$40} is one of the most studied ones, with observations available along the entire electromagnetic spectrum. The object is a binary system located at $(l, b) = (344\fdg98, 2\fdg45)$, 
which contains a 5-7 $M_{\sun}$ black hole with a companion star of 2.3 $M_{\sun}$ (Bailyn et al. \cite{bailyn95}). Its origin has been discussed in the past in different frameworks. \cite{israelian99} detected large overabundances of $O$, $Mg$, $Si$ and $S$ in the atmosphere of the companion star, which were interpreted as evidence for supernova ejecta captured by the stellar atmosphere. The relative abundances of these elements suggest that the supernova progenitor could be a massive star with 25 to 40 solar masses. 

Using radio continuum and HI-line observations around the previously accepted distance of 3.2 kpc for \object{GRO~J1655$-$40} (which corresponds to a velocity range from -42 to -30 km s$^{-1}$), \cite{combi01} found a cavity with "horse-shoe" morphology in the large-scale HI distribution. This kind of structure has been already observed towards early-type stars with very strong stellar winds (Benaglia \& Cappa \cite{benaglia99}) and supernova remnants (Butt et al. \cite{butt01}).

Other astrophysical connection between \object{GRO~J1655$-$40} and its environment has been suggested by \cite{mira02}. These authors have used Hubble Space Telescope measurements to compute a runaway space velocity of 112$\pm$18 km s$^{-1}$ for the object, which is moving in opposite direction of the open cluster NGC 6242, being the latter located at $(l, b) = (345\fdg47, 2\fdg45)$. According to photometric and radial velocity measurements (Glushkova et al. \cite{glus97}) this open cluster is located at a distance of 1.02$\pm$0.1 kpc from the Sun. Therefore, such connection could not be confirmed since the old estimate of the distance for the microquasar had a value different from the one of the open cluster. Recently, the distance to \object{GRO~J1655$-$40} has been revisited by \cite{foell06a}. Using ESO archive and new VLT-UVES spectra they have determined a spectral type F6IV for the secondary star of the binary system. Thus, the distance to \object{GRO~J1655$-$40} would be smaller than 1.7 kpc. More recently, \cite{foell06} showed that although this upper limit is rather firm, the lower limit is not well-constrained. 
Therefore, \object{GRO~J1655$-$40} could be one of the closest known black holes to the Sun.

In a recent paper on \object{GRO~J1655$-$40} by \cite{sala07} using XMM data the authors derive a link between the inner radius of the accretion disk and the distance. They argue that if the distance were smaller than 1.7 kpc and the mass is 5 $M_{\sun}$, then the inner accretion disk radius will be inside the gravitational radius of the black hole. However, the mass and spin of the black hole are uncertain. If the black hole were rapidly rotating as suggested by \cite{McClintock} and its mass less than 5 $M_{\sun}$, the horizon of the Kerr black hole could be well inside the inner accretion radius. Furthermore, the authors admite that an estimate of the inner accretion radius using XMM data is model dependent. 

Motivated by the new determination of the distance to the microquasar, we conduct here a detailed study of the distribution of the atomic (HI) and molecular (CO) gas, as well as of the dust in the vicinity of the open cluster NGC 6242. Our main goal is to detect the traces in the interstellar medium (ISM) around the cluster (i.e., a cavity), left by some explosive event. The proximity and relative height of the cluster on the galactic disk provides a unique opportunity to study SNR evolution and the possible physical link with microquasar formation. The paper is organized as follows: in Sect.~\ref{observations} we describe our HI and CO observations. In Sect.~\ref{results} we present the main results and then we discuss the nature of the detected structures, and their possible physical connections with \object{GRO~J1655$-$40}, in Sect.~\ref{discussion}.

\section{HI and CO observations} \label{observations}

\subsection{HI observations} 

We have performed HI observations towards \object{GRO~J1655$-$40} 
with a 30-m single dish telescope located at the Instituto
Argentino de Radioastronom\'{\i}a (IAR), Villa Elisa, Argentina.
The observations were carried out during six consecutive sessions
on November 20-25, 2004. The radiotelescope has a helium-cooled HEMT
amplifier with a 1008-channel autocorrelator, attaining a system temperature 
$T_{\rm sys} \approx$ 35 K. System parameters and additional details 
of the observational technique can be found in \cite{combi98}. The HI line was
observed in hybrid total power mode and the sky was sampled on a
$0.35^{\circ}$ rectangular grid. Each grid position was observed
during 60~s with a velocity resolution of \mbox{$\sim 1$ km
s$^{-1}$} and a coverage of $\pm 450$ km s$^{-1}$. A set of HI
brightness temperature maps ($\Delta T_{\mbox{\scriptsize rms}}
\sim 0.2$ K) were made for the velocity interval $\Delta v=$($-80$
km s$^{-1}$, $+20$ km s$^{-1}$). The data were analyzed using AIPS 
routine programs. 

\begin{figure}[t!] 
\center
\resizebox{\hsize}{!}{\includegraphics[angle=0]{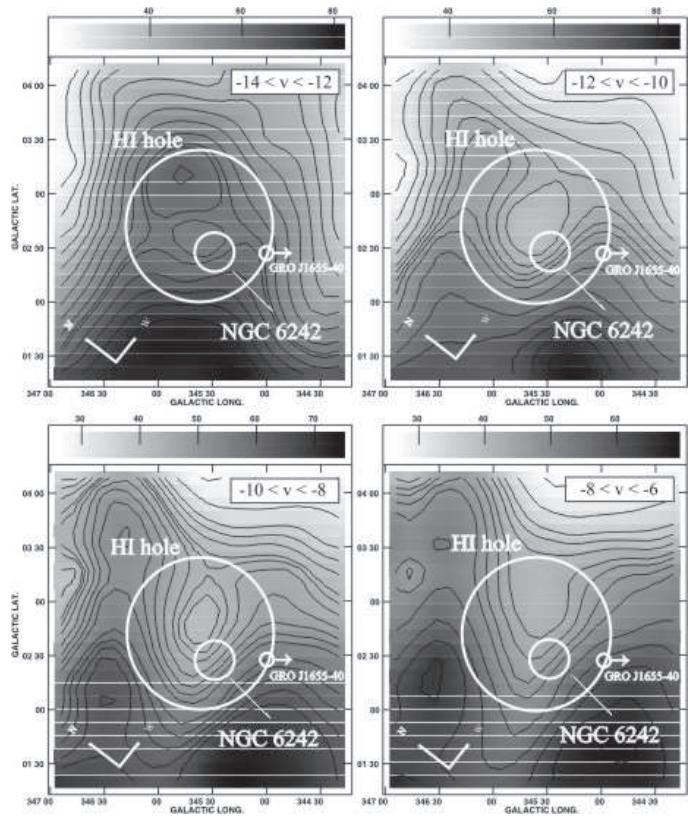}}
\caption{HI brightness temperature channel maps obtained for the velocity range from $-14$ to $-6$ km s$^{-1}$ around the
position of NGC 6242 (indicated with a white small circle). In all images the contours are labeled in steps of 4 K, starting from 4 K. 
A J2000 frame is shown for reference in the figure and all images. The location and motion direction of \object{GRO~J1655$-$40} are shown with a small circle and an arrow, respectively.}
\label{fig:h1}
\end{figure}

\subsection{CO observations} 

The $^{12}$CO data were observed during 2002, October with the 4 m NANTEN telescope of Nogoya University, located then at Las Campanas (Carnegie Institution of Washington) in Chile. At the $^{12}$CO ($J$=1-0) transition of 115.27 GHz, the angular resolution (HPBW) is 2'.8. A field of 1$^{\circ}$.45 $\times$ 1$^{\circ}$.45 centered on $l$=345$^{\circ}$.45, $b$=+2$^{\circ}$.5 was observed in the $^{12}$CO line with an integration time of 40 s for each position. The typical rms noise antenna temperature of the observations is in the range of 0.2-0.25 K, and the velocity resolution is of 0.055 km s$^{-1}$. 

The data were reduced with the NANTEN data reduction software. Baselines were fitted to each individual spectrum. The resulting data cubes were transformed into FITS format, and further processing was done using the AIPS software. 

\begin{figure}[t!] 
\center
\resizebox{\hsize}{!}{\includegraphics[angle=0]{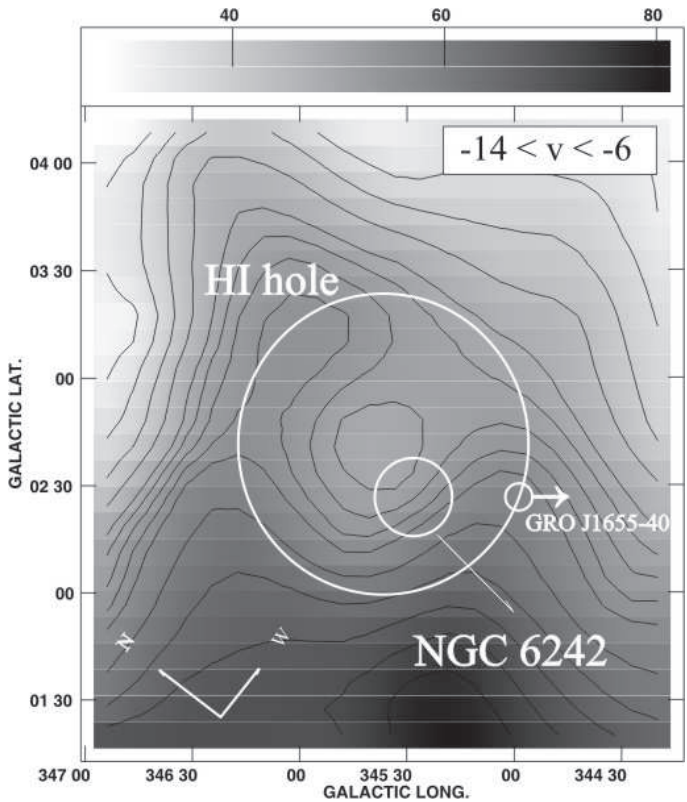}}
\caption{Integrated column density map obtained
for the velocity range from $-14$ to $-6$ km s$^{-1}$ around the
position of NGC 6242 (indicated with a white circle). The contours are labeled in steps of 4 K, starting from 4 K. The location and motion direction of \object{GRO~J1655$-$40} are shown with a small circle and an arrow, respectively.}
\label{fig:co}
\end{figure}

\begin{figure}[t!] 
\center
\resizebox{0.8\hsize}{!}{\includegraphics[angle=0]{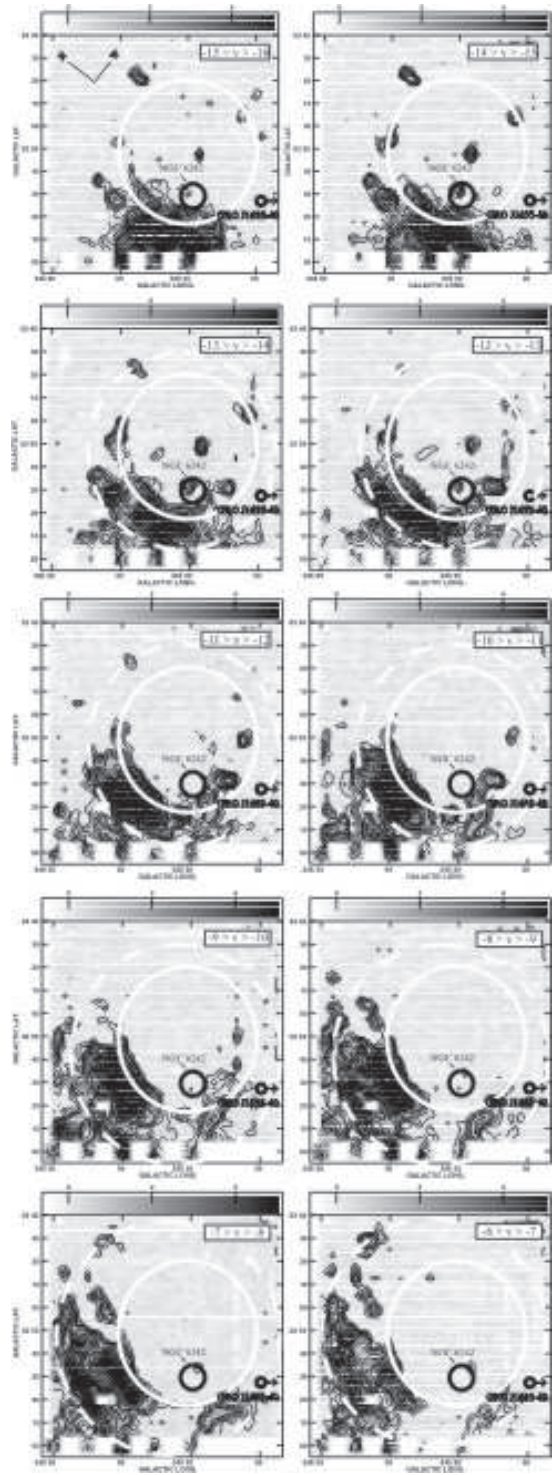}}
\caption{The gray-scale channels map indicate the intensity of CO($J$=1$\rightarrow$0) emission of the ambient molecular material, in the LSR velocity interval from $-16$ to $-6$ km s$^{-1}$ around the position of NGC 6242 (indicated with a black circle). White dashed circles show the expansion of the CO material along the velocity range. The location and motion direction of \object{GRO~J1655$-$40} are shown with a small circle and an arrow, respectively.}
\label{fig:co}
\end{figure}

\section{Main Results} \label{results}

If the physical link between \object{GRO~J1655$-$40} and NGC 6242 is real, the system should have been ejected from the cluster $\sim$ 3.4$\times$ 10$^{5}$ years ago. After this time the radio emission of a SNR could be very faint or not detectable. In order to confirm this hypothesis, we search for signatures of some explosive  event around NGC 6242 at radio frequencies using data from the 4.85-GHz (Condon et al. \cite{condon93}) and 2.7-GHz surveys (Duncan et al. \cite{duncan95}), respectively. As expected, we found no traces of any SNR phenomena towards the region. 

\subsection{Atomic component}

Standard galactic rotation models (Russeil \cite{russeil03}), corrected by excess associated with part of the Carina arm (Alvarez et al. \cite{alvarez90}), indicate that for a kinematic distance in the range of 0.8 to 1.6 kpc (the minimum distance to NGC 6242 and the maximun distance to \object{GRO~J1655$-$40}), corresponds a velocity interval of $-14$ to $-6$ km s$^{-1}$. Fig.1, displays HI brightness temperature channel maps of the relevant interval. The development of a cavity centered at $(l,b)$=(345.67$^{\circ}$,+2.68$^{\circ}$) can be clearly seen in these maps. The angular size of this structure is $\sim$ 1$^{\circ}$.5 $\times$ 1$^{\circ}$.5. Thus, at an average distance of 1.2 kpc, the linear size of the HI structure would be $\sim$ 32 pc. The location of NGC 6242 and the HI cavitiy are indicated with white circles. 

The angular extension of the cavity increases from -14 to -10 km s$^{-1}$. Between -12 to -10 km s$^{-1}$, the hole reaches its maximun size and forms a kind of "horse-shoe" structure at velocities larger than -8 km s$^{-1}$. The HI map integrated over the velocity interval from $-14$ to $-6$ km s$^{-1}$ is shown in Fig.2. 

Given that, no radio emission was detected towards the HI hole, we have computed the upper-limit for the radio surface brightness emission at 1 GHz. Flux densities at 2.7 and 4.8 GHz of the region covered by the HI hole and an angular size for the cavity of 1.5$^{\circ}$ were used. As may be expected, it turns out to be very low, with $\Sigma_{\rm 1 GHz} \sim$ 4($\pm$2)$\times$10$^{-24}$  W m$^{-2}$ Hz$^{-1}$ sr$^{-1}$.
 
\begin{figure}[t!] 
\center
\resizebox{1.0\hsize}{!}{\includegraphics[angle=0]{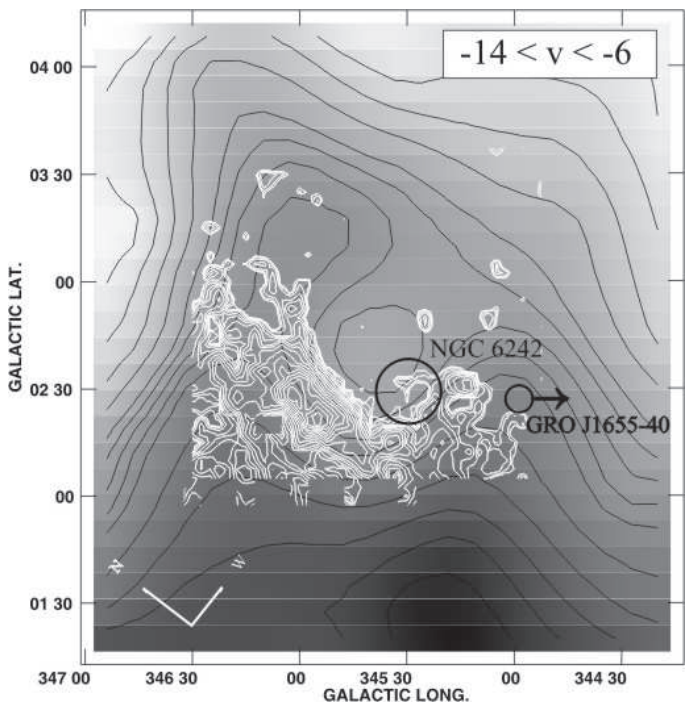}}
\caption{Comparison of the HI distribution (in grey-scale) in the velocity range from $-14$ to $-6$ km s$^{-1}$ with the 
$^{12}$CO emission integrated in the same velocity range. In the HI image the contour are labeled in steps of 4 K, starting from 4 K.  
The plotted $^{12}$CO contours (white lines) are 2, 2.5, 3, 3.5, 6.5, 9.5, 16.5, 25.5 and 35.5 K km s$^{-1}$. The location and motion direction of \object{GRO~J1655$-$40} are shown with a small circle and an arrow, respectively.}
\label{fig:integra}
\end{figure}

\subsection{Molecular component}

Figure 3 shows the distribution of the CO($J$=1$\rightarrow$0) emission in a smaller area around NGC 6242 (indicated with a black circle). A large amount of compressed CO lies along the lower half of the image, in the region closer to the Galactic plane. This extended molecular feature coincides with the lower border of the HI cavity. It has a semi-circular morphology, typical of the interaction of a SNR shock with the steep density gradient towards the galactic plane. From -16 to -6 km s$^{-1}$ the angular extension of the cavity increases with the velocity. It is evident that in this velocity interval the molecular material was displaced to form part of the cavity seen in the HI distribution. 

At the south region of the cluster the CO distribution exhibits a peculiar bifurcation. The morphology of the material  could be the result from the eruption of a shock into a region of very low density. When the shock front hits a low density region it expands more rapidly into the interstellar medium producing a separation. The good agreement of the HI material, similar morphology and behavior of the molecular gas in this velocity range, suggests that all these structures have a physical link and originate in the same explosive event. 
 
A comparison of the HI distribution integrated in the velocity range from $-14$ to $-6$ km s$^{-1}$ with the $^{12}$CO emission integrated in the same velocity range, is shown in Fig. 4. It is interesting to note that the molecular material is mainly displaced towards the lower region of the HI cavity. This is consequence of the fact that the mean thickness of the molecular disk is less than that of atomic gas (Bronfman et al. \cite{bronfman88}). The circle indicate the position and aproximate extension of the open cluster NGC 6242.  

\subsection{Infrared emission}

We have also searched for traces of a SNR shock using {\it IRAS} and {\it MSX} observations. Figure 5 shows the MSX infrared emission at 8.28 $\mu$m around NGC 6242. The size of the image represents the HI hole shown in Fig.2. In spite that is a very confuse region of the Galaxy, we found that several IR structures with extended morphology have good correlation with the molecular emission along the southern part of the image. At the central region of the HI hole we found extended IR emission with a semi-circular morphology (feature A). For all these structures the IR emission follows the spectral trend F$_{12} \leq$ F$_{25} \leq$ F$_{60} \leq$ F$_{100}$ (where F$_{\lambda}$ is the flux density at wavelenght $\lambda$) characteristic of shocked-heated dust (Junkes et al. \cite{junkes92}). At the southern part of the image, we can also see a large amount of IR emission, which could represent the shocked dust component of the associated CO material.    

An increase in the 60/100 $\mu$m IR color is a useful tracer of shocked-heated dust. The 60/100 $\mu$m color-corrected ratio for the observed infrared emission of the whole region is shown in Fig.6. It is interesting to note, that the infrared ratio level changes abruptly on the south border of the HI hole, which suggests the presence of the SNR shock. The small black circle represents the aproximate extension of NGC 6242. Enhanced IR emission on the central region and the southern part of the HI hole, is revealed by the ratio 60/100 $\mu$m. This values agree very well with the color temperature derived for IRAS sources associated with shocked material in old SNRs (Huang et al. \cite{huang86}). 

\section{Discussion} \label{discussion}

\begin{figure}[t!] 
\center
\resizebox{1.0\hsize}{!}{\includegraphics[angle=0]{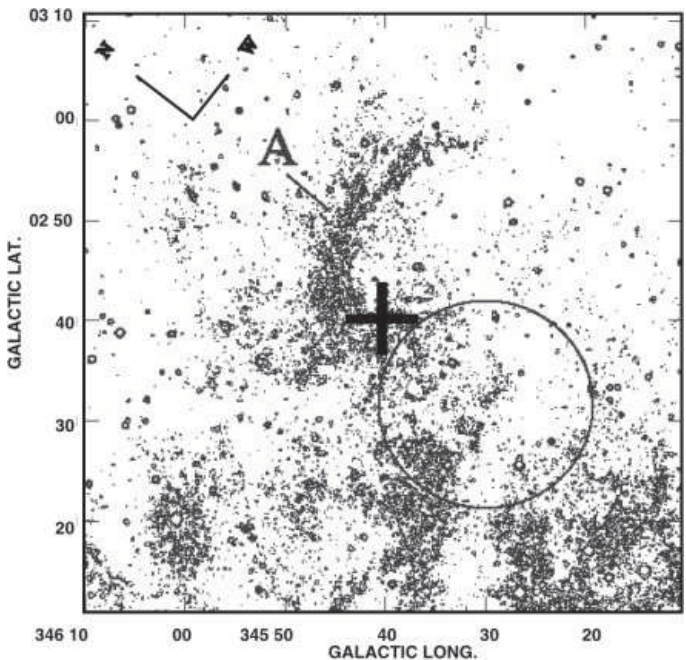}}
\caption{Infrared emission from MSX at 8.28 $\mu$m. The position of the NGC 6242 is indicated with a black circle. The location of the center of the HI hole is indicate with a black cross. The size of the figure is approximately that of the HI cavity. The central semi-circular feature is labeled with letter A.}
\label{fig:integra}
\end{figure}

Atomic and molecular observations reveal that NGC 6242 is surrounded by shocked material with semi-circular structure, typical of a medium where a large amount of energy has been release. The existence of a local minimum in the HI distribution at the approximated position of the cluster is evident. It is confirmed by the CO distribution which is bending on the lower border of the HI cavity. Moreover, infrared emission found within and along the south-eastern border of the HI hole, has properties typical of dust heated by collisions with the postshock gas. All these facts suggest that the detected emission has an  underlying physical connection originated in the same explosive phenomenon. 

\begin{figure}[t!] 
\center
\resizebox{1.0\hsize}{!}{\includegraphics[angle=0]{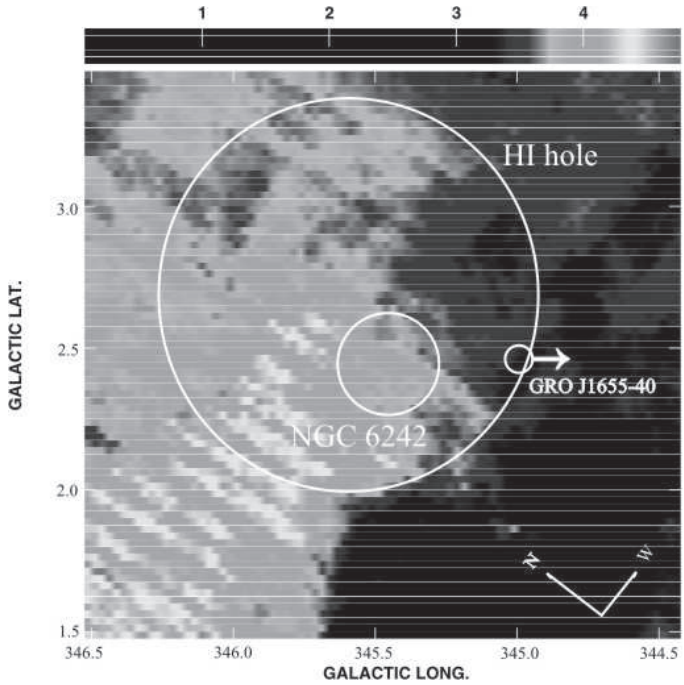}}
\caption{Gray-scale distribution of the 60/100 $\mu$m infrared ratio, which varies between 0.5 and 5. Note as the infrared ratio level changes  abruptly on the south border of the HI hole. The position of the HI hole and NGC 6242 are indicated with white circles. The location and motion direction of \object{GRO~J1655$-$40} are shown with a small circle and an arrow, respectively.}
\label{fig:integra}
\end{figure}

The energetic impact of a SN explosion that occurs out of the galactic disk can drive an asymmetric expanding bubble if the surrounded ambient gas is non-uniform. The shock-front could severely affect the dynamic structure of the surrounding gas, originating the observed semi-circular cavity, and heating the ambient dust. Assuming that the HI - CO cavity is the signature left by the action of a SNR on the surrounding ISM, we can use the N H I  images and CO maps to estimate the mass swept-up by the front-shock. It allows to obtain the initial ambient gas density and roughly to estimate the time elapsed since the SN explosion. 
  
Estimates of the physical parameters for the HI structure were derived by fitting a circle to the maximum of the envelope, which covers an area of $\Delta l \times \Delta b \approx$ 1$^{\circ}$.5 $\times$ 1$^{\circ}$.5. The integrated column densities shown in Fig.2 were used to estimate the total mass displaced to form the cavity. If the background density is given by the density at the minimum, we obtain that at least 9.000 $M_{\odot}$ have been removed from this region. The expansion velocity of the shell is defined approximately as ($\Delta$v/2)+2 km s$^{-1}$, where $\Delta$=v$_{2}$-v$_{1}$ is the velocity interval through which the cavity is seen. Using an average expansion velocity of $\sim$12 km s$^{-1}$, we obtain a lower limit to the original energy released in the explosion of $\sim$ 2$\times$10$^{49}$ erg. Since the SNR can have lost a significant part of its energy via radiative cooling, the initial energy could have been $\sim$ 10$^{51}$ erg. The mean density before the explosion (i.e. the initial density) can be obtained using the removed mass from the envelope's volume. Averaging the column densities within the cavity, we obtain that the density due to atomic material is $\sim$ 14 cm$^{-3}$. 

We further estimate the molecular component (H2) of the initial density, averaging 41 molecular clouds of the Carina arm taken from the study of \cite{grabelsky88}. Taking into account the region of the lower half of the HI hole and including the correction of He, we obtain that the density of  molecular hydrogen would be $\sim$ 14 cm$^{-3}$, that corresponds to 7500 $M_{\odot}$ (which renders a total atomic and molecular displaced mass of 16.500 $M_{\odot}$). Assuming that the whole cavity has been emptied of HI material and only the lower half of CO, we estimate that an upper limit for the initial density, averaged over the whole cavity, is N(H) $\sim$ 28 cm$^{-3}$. With this density and a radius of 16 pc for the cavity, we can estimate the lower limit for the SNR's age using the dynamical evolution equations for an expanding SNR, described by \cite{koo04}. The non-detection of radio emission and size of the SNR's radius, suggest that it is well within the radiative phase. The transition to this phase in SNR evolution is often triggered when the blast wave interacts with dense atomic and molecular clouds of gas in the ISM, a fact supported by the large amount of shocked gas observed in the region. 

According to \cite{koo04}, the maximum age of the HI hole due to the SNR 
is defined as the time when the shock velocity drops to the ambient sound speed, so that the interior pressure is comparable to the ambient pressure and the shell breaks up.
In this case, the lifetime of the SNR is described by:

\begin{eqnarray}
\tau = 2.02 \times 10^{6} (\beta c_{\rm s,6})^{-1.43} n_{0}^{-0.367} E_{51}^{0.316} \rm yr - 0.91 t_{\rm sf} 
\end{eqnarray}

where c$_{s,6}$=c$_{s}$/(10$^{6}$ cm s$^{-1}$), c$_{s}$ is the sound speed, $\beta$=2, E$_{51}$ is the SN energy released to the ISM in units of 10$^{51}$ erg, $n_{0}$ is the 
medium density, and $t_{sf}$ is the shell formation time given by,

\begin{eqnarray}
t_{sf} = 3.61 \times 10^{4} n_{0}^{-4/3} E_{51}^{3/14} \rm yr. 
\end{eqnarray}

Using a mean density of $n_{0} \sim$ 28 cm$^{-3}$ and a SN energy of $\sim$ 10$^{51}$ erg, we obtain $t_{sf}$= 400 yr. For c$_{s,6}$=1, the lower limit  to the lifetime of the SNR is $\tau$= 2.2$\times$10$^{5}$ yr.

This value is on the order of the time elapsed by \object{GRO~J1655$-$40} in moving from NGC 6242 up to its present position, which is $\sim$ 3.4$\times$ 10$^{5}$ yr.

\section{Summary} \label{summary}

We have investigated the surroundings of the open cluster NGC 6242 through HI, CO, and IR observations. Several facts corroborate the hypothesis of 
a SN event occurring within or very close to NGC 6242.
\begin{itemize}

\item The HI brightness temperature channel maps for the distance range of 1.2$\pm$0.4 kpc show the development of a cavity with an angular size of $\sim$ 1$^{\circ}$.5 $\times$ 1$^{\circ}$.5. 

\item The distribution of the CO($J$=1$-$0) emission displays a large amount of compressed material with semi-circular morphology along the south and east borders of the HI cavity. 

\item Infrared observations reveal the presence of several IR structures with extended morphology and good correlation with atomic and molecular emissions. The 60/100 $\mu$m color-corrected ratio shows also enhanced infrared emission towards the south-eastern part of the HI hole. These values agree very well with the color temperature derived for IRAS sources associated with shocked material in old SNRs. 

\item All the available information suggests that the detected emissions morphologies have a physical link, originated in a supernova explosion occurred within or very close to the cluster a $t$ $\ga$ 2.2$\times$ 10$^{5}$ years. If the association between the open cluster and the SNR is confirmed then the distance of the microquasar would be better known.  

\end{itemize}

Now, is the origin of \object{GRO~J1655$-$40} physically associated to the explosive event?. We suggest that a possible scenario, which fits the data presented in this work, is that an old supernova explosion occurred at least 2.2$\times$ 10$^{5}$ years ago inside NGC 6242. In this explosive event \object{GRO~J1655$-$40} could originate and consequently be ejected up to its present position in the sky. A similar case, not confirmed, could be that of the microquasar LS 5039 and the SNR G016.8-01.1 (\cite{ribo02}).
  
Unveiling the origin of the microquasar will allow to constrain its distance. If the present evidence in this work 
is confirmed, \object{GRO~J1655$-$40} would be the first microquasar physically linked to a very dilute, but still detectable SNR.

High-resolution HI observations are required for a most detailed study of the cold gas and evidence of cloud compression originated by the shock front.  Investigations of other molecular species, such as OH, HCO+ and H2CO, could help in search of shock signatures, particularly around the south-eastern region of the HI hole, where a large amount of gas and dust is detected.
  
\begin{acknowledgements}
We are very grateful to the referee for detailed suggestions which improved the paper significantly.
We thank G.E. Romero and J. Mart\'{\i} for their critical reading of the manuscript. Remarks by T. Mu\~noz Darias are also acknowledge.
J.A.C. is a researcher of the programme {\em Ram\'on y Cajal} funded jointly
by the Spanish Ministerio de Educaci\'on y Ciencia (former Ministerio de
Ciencia y Tecnolog\'{\i}a) and Universidad de Ja\'en.
The authors also acknowledge support by DGI of the Spanish Ministerio de
Educaci\'on y Ciencia under grants AYA2004-07171-C02-02 and
AYA2004-07171-C02-01, FEDER funds and Plan Andaluz de Investigaci\'on of Junta
de Andaluc\'{\i}a as research group FQM322. This research made use of data products from the Midcourse Space Experiment. Processing of the data was funded by the Ballistic Missile Defense Organization with additional support from NASA 
Office of Space Science. This research has also made use of the NASA/ IPAC Infrared Science Archive, which is operated by the Jet Propulsion Laboratory, California Institute of Technology, under contract with the National Aeronautics and Space Administration. L.B. acknowledges support from the Chilean Center for Astrophysics FONDAP 15010003     
         
\end{acknowledgements}


\begin{thebibliography}{}

\bibitem[1990]{alvarez90} Alvarez, H., May, J., 
\& Bronfman, L.\ 1990, \apj, 348, 495

\bibitem[1995]{bailyn95} Bailyn, C.~D., Orosz, 
J.~A., McClintock, J.~E., \& Remillard, R.~A.\ 1995, \nat, 378, 157 

\bibitem[1999]{benaglia99} Benaglia, P., \& 
Cappa, C.~E.\ 1999, \aap, 346, 979 

\bibitem[1988]{bronfman88} Bronfman, L., Cohen, 
R.~S., Alvarez, H., May, J., \& Thaddeus, P.\ 1988, \apj, 324, 248 


\bibitem[2001]{butt01} Butt, Y.~M., Torres, 
D.~F., Combi, J.~A., Dame, T., \& Romero, G.~E.\ 2001, \apjl, 562, L167 

\bibitem[Combi et al. (1998)]{combi98} Combi, J.~A., Romero, 
G.~E., \& Arnal, E.~M.\ 1998, \aap, 333, 298 

\bibitem[Combi et al. (2001)]{combi01} Combi, J.~A., Romero, 
G.~E., Benaglia, P., \& Mirabel, I.~F.\ 2001, \aap, 370, L5 

\bibitem[1993]{condon93} Condon, J.~J., Griffith, 
M.~R., \& Wright, A.~E.\ 1993, \aj, 106, 1095 

\bibitem[1995]{duncan95} Duncan, A.~R., Stewart, 
R.~T., Haynes, R.~F., \& Jones, K.~L.\ 1995, \mnras, 277, 36 

\bibitem[Foellmi et al. (2006)]{foell06a} Foellmi, C., Depagne, 
E., Dall, T.~H., \& Mirabel, I.~F.\ 2006, \aap, 457, 249 

\bibitem[Foellmi (2006)]{foell06} Foellmi, C., proc. of "The Future of Photometric, 
Spectrophotometric and Polarimetric Standardization", ed.: C. Sterken, ASP Conf. Series. 2006

\bibitem[2006]{glus97} Glushkova, E.~V., 
Zabolotskikh, M.~V., Rastorguev, A.~S., Uglova, I.~M., \& Fedorova, A.~A.\ 
1997, Astronomy Letters, 23, 71 

\bibitem[Grabelsky et al. (1987)]{grabelsky87} Grabelsky, D.A., Cohen, R.S., 
Bronfman, L., Thaddeus, P., \& May, J. 1987, \apj, 315, 122

\bibitem[Grabelsky et al. (1988)]{grabelsky88} Grabelsky, D.~A., 
Cohen, R.~S., Bronfman, L., \& Thaddeus, P.\ 1988, \apj, 331, 181 

\bibitem[1992]{junkes92} Junkes, N., Fuerst, E., 
\& Reich, W.\ 1992, \aap, 261, 289 

\bibitem[1986]{huang86} Huang, Y.-L., Dickman, 
R.~L., \& Snell, R.~L.\ 1986, \apjl, 302, L63 

\bibitem[Israelian et al. (1999)]{israelian99} Israelian, G., 
Rebolo, R., Basri, G., Casares, J., \& Mart{\'{\i}}n, E.~L.\ 1999, \nat, 
401, 142 

\bibitem[Koo \& Kang (2004)]{koo04} Koo, B.-C., \& Kang, 
J.-h.\ 2004, \mnras, 349, 983 

\bibitem[McClintock et al. 2006]{McClintock} McClintock, J. ~E.et al. \ 2006, \apj, 652, 518 

\bibitem[1999]{mira99} Mirabel, 
I.~F., \& Rodr{\'{\i}}guez, L.~F.\ 1999, \araa, 37, 409 

\bibitem[Mirabel et al. (2002)]{mira02} Mirabel, I.~F., 
Mignani, R., Rodrigues, I., Combi, J.~A., Rodr{\'{\i}}guez, L.~F., \& 
Guglielmetti, F.\ 2002, \aap, 395, 595 

\bibitem[2004]{mira04} Mirabel, I.~F.\ 2004, Revista 
Mexicana de Astronomia y Astrofisica Conference Series, 20, 14 

\bibitem[2005]{paredes05} Paredes, J.~M.\ 2005, Chinese 
Journal of Astronomy and Astrophysics, 5, 121 

\bibitem[Rib\'o et al. 2002]{ribo02} Rib{\'o}, M., Paredes, 
J.~M., Romero, G.~E., Benaglia, P., Mart{\'{\i}}, J., Fors, O., \& 
Garc{\'{\i}}a-S{\'a}nchez, J.\ 2002, \aap, 384, 954 

\bibitem[2003]{russeil03} Russeil, D. 2003, \aap, 397, 133 

\bibitem[Sala et al. (2007)]{sala07} Sala, G., Greiner, J., 
Vink, J., Haberl, F., Kendziorra, E., \& Zhang, X.~L.\ 2007, \aap, 461, 
1049
 
\end{thebibliography}
\end{document}